\documentclass[journal,twoside,web]{ieeecolor}
\usepackage{generic}
\usepackage{cite}
\usepackage{amsmath,amssymb,amsfonts}
\usepackage{algorithmic}
\usepackage{graphicx}
\usepackage{algorithm,algorithmic}
\usepackage{hyperref}
\hypersetup{hidelinks=true}
\usepackage{textcomp}
\usepackage{orcidlink}
\usepackage{longtable}
\usepackage{booktabs}
\usepackage{tabularx}
\usepackage{makecell}

\usepackage{listings}

\def\BibTeX{{\rm B\kern-.05em{\sc i\kern-.025em b}\kern-.08em
    T\kern-.1667em\lower.7ex\hbox{E}\kern-.125emX}}
\begin{document}
\title{OMOP ETL Framework for Semi-Structured Health Data}
\author{Jacob Desmond\,\orcidlink{0009-0006-9755-1133},Ryan Wartmann, Chng Wei Lau\,\orcidlink{0000-0002-9652-134X}, Steven Thomas\,\orcidlink{0000-0002-2416-0020}, Paul M. Middleton\,\orcidlink{0000-0003-0760-1098}, Jeewani Anupama Ginige\orcidlink{0000-0002-6695-6983}
    \thanks{Jacob Desmond, Chng Wei Lau and Jeewani
Anupama Ginige are with School of Computer, Data and Mathematical Sciences, Western Sydney University, Penrith, NSW 2751 Australia (email: J.Desmond@westernsydney.edu.au; c.lau@westernsydney.edu.au; j.Ginige@westernsydney.edu.au). }
    \thanks{Ryan Wartmann is an independent research (e-mail: ryanwartmann1998@gmail.com). }
    \thanks{Steven Thomas and and Paul M. Middleton are with South Western Emergency Research Institute, Ingham Institute of Applied MedicalResearch, 1 Campbell St, Liverpool, 2170, NSW, Australia. (e-mail: Steven.Thomas@health.nsw.gov.au; Paul.Middleton@health.nsw.gov.au). }}

\maketitle

\begin{abstract}
    Healthcare data are generated in many different formats, which makes it difficult to integrate and reuse across institutions and studies. Standardisation is required to enable consistent large-scale analysis. The OMOP-CDM, developed by the OHDSI community, provides one widely adopted standard. \\
    Our framework achieves schema-agnostic transformation by extending upon existing literature in using human-readable YAML specification to support both relational (Microsoft SQL Server (MSSQL)) and document-based (MongoDB) data sources.  It also incorporates critical production readiness features: provenance-aware mapping and support for incremental updates. \\
    We validated the pipeline using 2.7 million patient records and 27 million encounters across six hospitals spanning two decades of records. The resulting OMOP-CDM dataset demonstrated an acceptable level of data quality with a 97\% overall passing rate based on the OHDSI Data Quality Dashboard check.  Our work provides a reusable blueprint for large-scale data harmonisation, directly supporting real-world medical data research.  
\end{abstract}


\section{Introduction}
\label{sec:introduction}
\IEEEPARstart{T}{he} Observational Medical Outcomes Partnership (OMOP) Common Data Model (OMOP-CDM) standardises healthcare data to enable consistent analysis across institutions \cite{OHDSIwebsite,OHDSI_Athena,OHDSI_DataStandardization}. Maintained by OHDSI, it comprises ~40 clinical and vocabulary tables supporting interoperability in research.

The South Western Emergency Research Institute (SWERI) is adopting OMOP-CDM to improve access to clinical data. Its two-stage project first consolidates disparate systems into the CEDRIC repository \cite{SWSLHD_Fujitsu_CEDRIC}, then maps CEDRIC to OMOP-CDM.

This paper outlines transforming CEDRIC data, covering 2.7M patients and 27M episodes from six hospitals over 20 years, into OMOP-CDM. The resulting internal NSW Health database serves as a precursor to a fully anonymised public dataset. We review related work, describe CEDRIC and OMOP structures, and present the ETL approach used to preserve source fidelity under OMOP conventions, developed without direct access to the target database. We report outcomes and lessons for similar projects.

Finally, we frame research questions on configurability, automation and scalability of OMOP-CDM conformance, and handling semi-structured document-store data while preserving hierarchy.

This paper aims to address the following research questions:
\begin{enumerate}
    \item What is the feasibility of a configurable ETL pipeline for transforming arbitrary healthcare data into the OMOP-CDM? \label{rq1}
    \item How can such a pipeline facilitate end-to-end automation and incremental updates while supporting scalability to large datasets and balancing source data fidelity with adherence to OMOP-CDM standards? \label{rq2}
    \item What approaches can enable an ETL pipeline to transform healthcare data from heterogeneous repositories, ranging from structured relational databases (e.g., Microsoft SQL Server (MSSQL)) to semi-structured document stores (e.g., MongoDB), into the OMOP-CDM while preserving hierarchical information and ensuring semantic and structural consistency with the relational OMOP-CDM schemas? \label{rq3}
\end{enumerate}

The code and accompanying documentation for this work are publicly available in our GitHub repository \cite{self_repo}.

\section{Literature Review/Background/Similar Work}
\label{sec:background}
\subsection{What is OMOP?}
\label{subsec:WhatIsOMOP}
The OMOP-CDM is a standardised schema enabling harmonisation and secondary use of heterogeneous healthcare data (e.g., EHRs, claims, registries) for large-scale, federated research \cite{OHDSIwebsite}. By converting local data into a unified tabular structure with standard vocabularies (e.g., SNOMED CT, RxNorm, LOINC), it supports reproducible analyses for real-world evidence generation, including comparative effectiveness and pharmaco­vigilance studies.

OMOP-CDM comprises interrelated tables for core clinical entities: \texttt{PERSON} (demographics), \texttt{CONDITION\_OCCURRENCE} (diagnoses), \texttt{DRUG\_EXPOSURE} (medications), \texttt{MEASUREMENT} (labs), and \texttt{VISIT\_OCCURRENCE} (encounters), all linked via standard concept IDs. Additional tables capture procedures, observations, devices, and episode context; \texttt{OBSERVATION\_PERIOD} and \texttt{METADATA} record provenance and timing; \texttt{SOURCE\_TO\_CONCEPT\_MAP} supports traceability; and \texttt{NOTE}/\texttt{NOTE\_NLP} enable use of unstructured clinical text. This modular design supports scalable integration and analytic consistency across distributed research networks \cite{Unimelb_HaBIC_OMOP_CDM}.

\subsection{What is CEDRIC?}
\label{subsec:WhatIsCedric}
\begin{figure}[!t]
\includegraphics[width=\columnwidth]{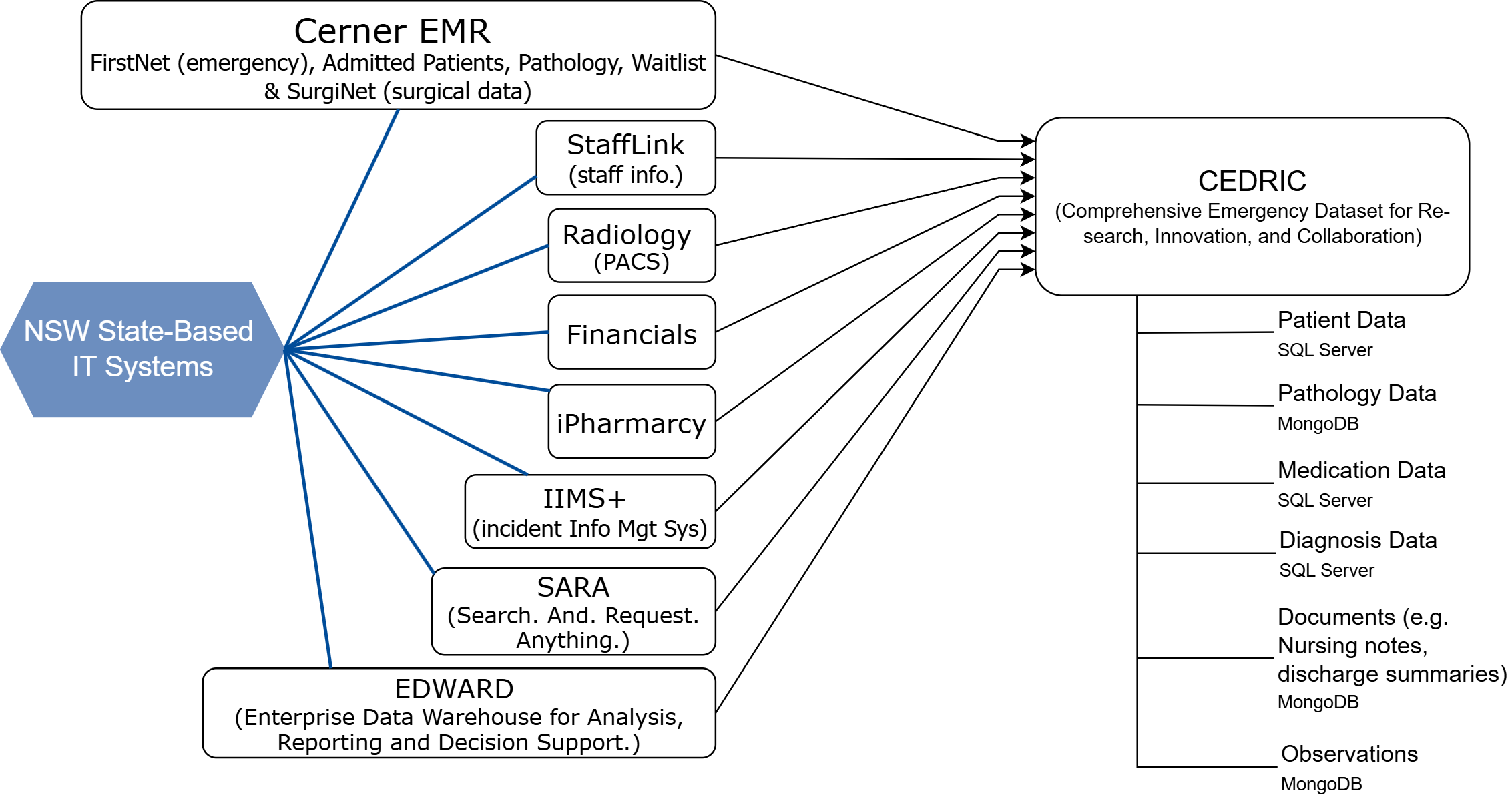}
\caption{Overview of CEDRIC}
\label{fig:cedric_overview}
\end{figure}

CEDRIC (Comprehensive Emergency Dataset for Research, Innovation and Collaboration) is a consolidated repository integrating data from NSW operational systems (Fig. \ref{fig:cedric_overview}). Feeds from Cerner EMR (FirstNet, admitted patients, pathology, SurgiNet), StaffLink, GE-PACS (radiology), financial/insurance systems, iPharmacy, IIMS+, SARA, and EDWARD converge into harmonised research domains such as demographics, pathology, medications, diagnoses, and clinical documents. CEDRIC stores structured data in MSSQL and document content in MongoDB, hosted within the secure health network.

\subsection{Background and Similar Work}
OMOP-CDM has become the leading standard for harmonising heterogeneous healthcare data. Numerous projects have transformed primary care, hospital EMRs, ICU, rare disease registries, and population cohorts into OMOP-CDM, enabling scalable cross-site analytics \cite{info:doi/10.2196/49542, info:doi/10.2196/30970, biedermann2021standardizing, 10.1093/jamiaopen/ooab001, 10.1093/jamia/ocac203, Tan2022OMOPBenefitRisk}.

Existing literature covers a large range of transformation effort:
\begin{enumerate}
    \item National-level EHR mapping - German, Norway, Estonia, and UK Biobank \cite{10.1093/jamiaopen/ooab001, 10.1093/jamia/ocac203, 10.1371/journal.pone.0311511, TRINH2024105602}.
    \item Rare disease registries - Rare disesase and pulmonary hypertension \cite{biedermann2021standardizing, 10.1093/jamiaopen/ooab001, 10.1093/jamia/ocac203, Tan2022OMOPBenefitRisk, wang2025scoping, kim2021transforming, YU2022104002, PENG2023104925, 10.1371/journal.pone.0311511, TRINH2024105602, 10.3389/fdata.2024.1435510}.
    \item Semi-structured / FHIR/ Novel data types - German FHIR data, YAML based ETL and PSG (waveform) + EHR \cite{kim2021transforming, PENG2023104925, UNSW_ETL}.
    \item COVID-19 focused - 8 countries COVID-19 data, and US PCORnet \cite{YU2022104002}.
    \item Other domain - Primary care, Benefit Risk assessment, MIMIC IIi ICU, informatic research , oncology and mental health \cite{info:doi/10.2196/49542, info:doi/10.2196/30970, Tan2022OMOPBenefitRisk, wang2025scoping, 10.3389/fdata.2024.1435510, ESPINOZA2023100119}.
\end{enumerate}

Most OMOP-CDM transformations use bespoke or semi-automated ETL pipelines, commonly leveraging OHDSI tools \cite{ohdsi_tools} alongside custom code in SQL, Python, R, Java, or ETL platforms \cite{info:doi/10.2196/49542, biedermann2021standardizing, 10.1093/jamiaopen/ooab001, 10.1093/jamia/ocac203, wang2025scoping, kim2021transforming, YU2022104002, PENG2023104925, 10.1371/journal.pone.0311511, TRINH2024105602, 10.3389/fdata.2024.1435510}. However, strict standard conformance can risk data loss when source structures or vocabularies misalign; mitigation usually involves custom concepts or minor schema adaptations while preserving interoperability \cite{biedermann2021standardizing, 10.1093/jamia/ocac203}.

Recent work explores more reusable, declarative approaches, e.g. configuration-driven frameworks using YAML to generate SQL for clearer and maintainable mappings \cite{Tan2022OMOPBenefitRisk, UNSW_ETL}. Remaining challenges include limited mapping automation, fidelity trade-offs, infrastructure-specific pipelines, and evolving schemas.

These gaps motivate our approach, addressed through RQ\ref{rq1}, RQ\ref{rq2}, and RQ\ref{rq3}.

\section{Methods, Techniques and Tools}
\label{sec:methods}

\subsection{Overall Approach}
This project was a collaboration between the South Western E-Health Research Institute (SWERI) and the School of Computer, Data and Mathematical Sciences (CDMS) at Western Sydney University, aiming to provide researchers with clinically relevant data for data-driven research.

CDMS developed a semi-automated framework to transform CEDRIC data (section~\ref{subsec:WhatIsCedric}) into OMOP-CDM (section~\ref{subsec:WhatIsOMOP}), despite not having direct access to the clinical dataset. Work proceeded using CEDRIC schema information and public OMOP-CDM documentation \cite{OHDSI_DataStandardization}. Closely informed by \cite{UNSW_ETL}, CDMS created YAML mappings (section~\ref{subsubsec:YAMLMappingFiles}) and a Python ETL framework to populate OMOP-CDM, released publicly \cite{self_repo}.

\begin{figure}[!t]
\includegraphics[width=\columnwidth]{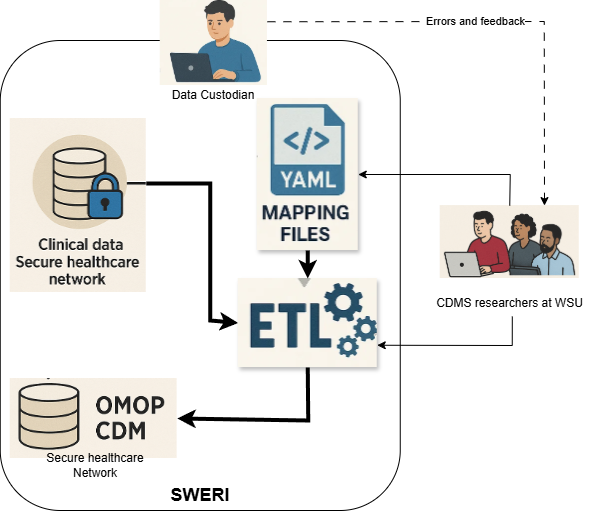}
\caption{Overall Process}
\label{fig:overall_process}
\end{figure}

The SWERI data custodians executed the transformation within the secure health network. Errors identified during the process (Fig.~\ref{fig:overall_process}) were relayed back to CDMS for updates. This iterative workflow ensured data security and integrity while enabling accurate alignment with OMOP-CDM.

\subsection{Mapping from CEDRIC to OMOP}
\subsubsection{Python ETL Framework}
The ETL framework is implemented in Python using SQLAlchemy \cite{sqlalchemy} for dynamic query generation and transaction handling. All transformations load data into MS SQL as compiled SQL statements, while MongoDB sources are accessed via aggregation pipelines.

Each ETL job is configured in a YAML file, typically one per OMOP table, following \cite{UNSW_ETL}. We extend this format to support MongoDB and structured YAML transformations. The YAML specifies source and destination tables, primary key handling, and column-level logic (section~\ref{subsubsec:YAMLMappingFiles}).

A central mapping table, adapted from \cite{UNSW_ETL}, maintains relationships between source and OMOP primary keys for traceability (section~\ref{subsubsec:YAMLMappingFiles}).

For MS SQL sources, the framework compiles queries, validates table references, prepares the mapping table, inserts unmapped rows, and sequentially applies column updates. For MongoDB, data are first flattened via aggregation pipelines, staged into MS SQL, and then processed through the same mapping and insertion workflow (section~\ref{subsubsec:mongodb}).

\begin{figure}[!t]
\includegraphics[width=\columnwidth]{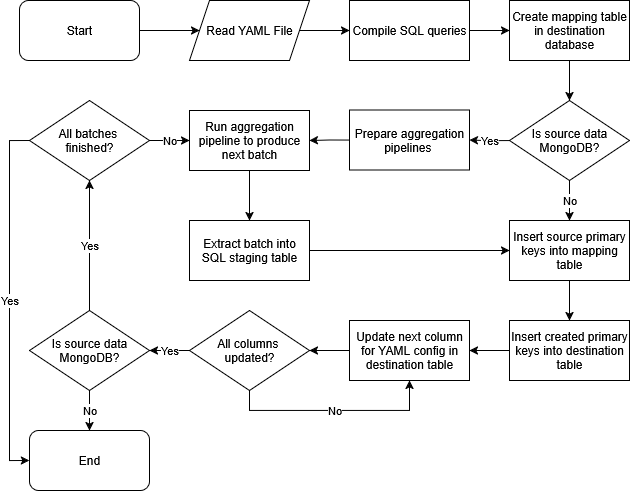}
\caption{High-level flowchart outlining key steps of the ETL process.}
\label{fig:etl_flow_chart}
\end{figure}

\subsubsection{YAML Mapping files}
\label{subsubsec:YAMLMappingFiles}
Each YAML file corresponds to a single OMOP-CDM destination table and defines the destination table name, source primary keys, and column-level mappings. These mappings are expressed as ordered transformation steps instead of raw SQL. We support five step types—\texttt{Column}, \texttt{Constant}, \texttt{StaticMapping}, \texttt{LookupMapping}, and \texttt{Func}—allowing minimal, composable logic (Table~\ref{tab:transform_types}). Following \cite{UNSW_ETL}, a per-table mapping table preserves links between source and generated OMOP primary keys, enabling incremental ETL.

We extend the mapping table with insert/update timestamps and a \texttt{processed} flag to support continuous updates, change tracking, and efficient reprocessing without duplication.

Where required, source references are replaced with aliased subqueries to avoid column conflicts, support joins (including self-joins), and enable multi-row generation (e.g., unpivoting attributes into multiple observations).

Most CEDRIC data are relational, but measurement data exist as nested MongoDB documents. To handle this, the YAML specification includes a \texttt{mongodb} section defining projection into a flat structure for staging before ETL execution (see section~\ref{subsubsec:mongodb}).

\subsubsection{MongoDB}
\label{subsubsec:mongodb}
To support ETL of the OMOP-CDM \texttt{MEASUREMENT} table, we developed a MongoDB extraction strategy that flattens nested documents into a relational format before loading. The flattening rules are fully defined in YAML and executed as a MongoDB aggregation pipeline, with output staged into an SQL table. Once staged, the existing ETL workflow proceeds unchanged.

Processing is performed in batches to avoid excessive in-memory buffering during document retrieval and staging. Batch control leverages the mapping table’s \texttt{processed} flag: only unprocessed source rows are extracted, and after each batch completes, a simple \texttt{UPDATE} marks those entries as processed. This allows scalable loading without modifying transformation logic, at the cost of one additional update per batch.

Flattening is configured using a YAML \texttt{fields} list, where each item defines:
\begin{itemize}
    \item the document field path to extract (using dot notation or list index syntax)
    \item the staging column name
    \item the SQL data type
\end{itemize}
Nested lists are handled by recursively defining additional \texttt{fields} blocks, allowing a single document to yield multiple rows. From this specification, the framework generates an aggregation pipeline that:
\begin{itemize}
    \item uses \texttt{\$unwind} to expand arrays
    \item applies \texttt{\$unionWith} to merge heterogeneous list paths
    \item uses \texttt{\$project} to enforce a consistent tabular schema
\end{itemize}
Missing attributes are represented as explicit nulls, enabling unified interpretation downstream. The \texttt{mongodb} YAML object is further described in Appendix~\ref{appendix:yaml} Table~\ref{tab:mongo_yaml}. An example of the flattening process can be seen in Figures~\ref{fig:mongodb_example_data}, \ref{fig:mongodb_example_yaml} and \ref{fig:mongodb_example_pipeline}, and Table~\ref{tab:mongodb_example_rows}.

Document and row-level filters can also be expressed directly in YAML as embedded MQL \texttt{\$match} objects, enabling pre and post-flattening restriction of measurement data. This ensures selective extraction of clinically relevant observations while maintaining a declarative, configuration-driven ETL process.

\begin{table}[!t]
\caption{Types of simple transformation steps used in YAML mappings}
\small
\setlength{\tabcolsep}{1mm}
\begin{tabular}{
    |>{\raggedright\arraybackslash}p{21mm}
    |>{\raggedright\arraybackslash}p{19mm}
    |>{\arraybackslash}m{42mm}
    |}
\hline
\textbf{Type} & 
\textbf{T-SQL Equivalent} & 
\textbf{Example Definition} \\
\hline
Column & 
\texttt{SELECT col FROM table} &
\begin{minipage}[t]{\linewidth}
\vspace{-0.5\baselineskip}
\begin{lstlisting}[basicstyle=\ttfamily\small,aboveskip=0pt,belowskip=0pt]
- type: Column
  table_name: |
    db.dbo.Patient
  column_name: Gender
\end{lstlisting}
\end{minipage} \\
\hline
Constant & 
A constant literal. &
\begin{minipage}[t]{\linewidth}
\vspace{-0.5\baselineskip}
\begin{lstlisting}[basicstyle=\ttfamily\small,aboveskip=0pt,belowskip=0pt]
- type: Constant
  value: 0
  type_str: INT
\end{lstlisting}
\end{minipage} \\
\hline
StaticMapping & 
\texttt{CASE} &
\begin{minipage}[t]{\linewidth}
\vspace{-0.5\baselineskip}
\begin{lstlisting}[basicstyle=\ttfamily\small,aboveskip=0pt,belowskip=0pt]
- type: StaticMapping
  map:
    Male: MALE
    Female: FEMALE
    Indeterminate: |
      AMBIGUOUS
  else: UNKNOWN
\end{lstlisting}
\end{minipage} \\
\hline
LookupMapping & 
\texttt{INNER JOIN} &
\begin{minipage}[t]{\linewidth}
\vspace{-0.5\baselineskip}
\begin{lstlisting}[basicstyle=\ttfamily\small,aboveskip=0pt,belowskip=0pt]
- type: LookupMapping
  table_name: |
    OMOP.dbo.CONCEPT
  mapto: concept_code
  retn: concept_id
  where: |
    OMOP.dbo.CONCEPT
      .domain_id
        = 'Gender'
\end{lstlisting}
\end{minipage} \\
\hline
Func & 
SQL Function execution. &
\begin{minipage}[t]{\linewidth}
\vspace{-0.5\baselineskip}
\begin{lstlisting}[basicstyle=\ttfamily\small,aboveskip=0pt,belowskip=0pt]
- type: Func
  name: extract
  args:
    - year
    - \value\
\end{lstlisting}
\end{minipage} \\
\hline
\end{tabular}
\label{tab:transform_types}
\end{table}

\begin{figure}[!t]
\begin{lstlisting}[frame=single,basicstyle=\ttfamily\small,breaklines=true]
[
  {
    "visit_id": 123456,
    "measures": [
      { "name": "systolic_bp" },
      { "name": "diastolic_bp" }
    ],
    "panels": [
      {
        "name": "fluid_intake",
        "measures": [
          { "name": "24hr" },
          { "name": "24hr_oral" }
        ]
      },
      {
        "name": "asthma_tracking",
        "measures": [
          { "name": "spells_per_week" },
          { "name": "max_expir_flow" }
        ]
      }
    ]
  },

  {
    "visit_id": 654321,
    "measures": [
      { "name": "systolic_bp" },
      { "name": "diastolic_bp" }
    ],
    "panels": [
      {
        "name": "fluid_intake",
        "measures": [
          { "name": "24hr" },
          { "name": "24hr_oral" }
        ]
      }
    ]
  }
]
\end{lstlisting}
\caption{An example of MongoDB documents which contain nested data to be flattened.}
\label{fig:mongodb_example_data}
\end{figure}

\begin{figure}[!t]
\begin{lstlisting}[frame=single,basicstyle=\ttfamily\small,breaklines=true]
mongodb:
  source:
    database: records
    collection: measures
  staging_table: OMOP.STAGING.MEASURES
  fields:
    - doc_key_path: visit_id
      staging_column: visit_id
      datatype: BIGINT

    - doc_key_path: measures
      fields:
        - doc_key_path: name
          staging_column: measure_name
          datatype: VARCHAR

    - doc_key_path: panels
      fields:
        - doc_key_path: name
          staging_column: panel_name
          datatype: VARCHAR

        - doc_key_path: measures
          fields:
            - doc_key_path: name
              staging_column: panel_measure_name
              datatype: VARCHAR
\end{lstlisting}
\caption{An example of the \texttt{mongodb} YAML object, specifying nested fields to be flattened in the example data shown in Fig.~\ref{fig:mongodb_example_data}}
\label{fig:mongodb_example_yaml}
\end{figure}

\begin{table}[!t]
\caption{Example Aggregation Pipeline Results}
\label{tab:mongodb_example_rows}
\setlength{\tabcolsep}{1mm}
\begin{tabular}{
    |p{10mm}
    |p{18mm}
    |p{22mm}
    |p{22mm}
    |}
\hline
\textbf{visit\_id} & \textbf{measure\_name} & \textbf{panel\_name} & \textbf{panel\_measure\_name} \\
\hline
123456 & systolic\_bp & NULL & NULL \\
123456 & diastolic\_bp & NULL & NULL \\
654321 & systolic\_bp & NULL & NULL \\
654321 & diastolic\_bp & NULL & NULL \\
123456 & NULL & fluid\_intake & 24hr \\
123456 & NULL & fluid\_intake & 24hr\_oral \\
123456 & NULL & asthma\_tracking & spells\_per\_week \\
123456 & NULL & asthma\_tracking & max\_expir\_flow \\
654321 & NULL & fluid\_intake & 24hr \\
654321 & NULL & fluid\_intake & 24hr\_oral \\
\hline
\multicolumn{4}{p{236pt}}{This table shows the flattened, row-like results given by running the example aggregation pipeline shown in Fig.~\ref{fig:mongodb_example_pipeline} on the example data shown in Fig.~\ref{fig:mongodb_example_data}.}\\
\end{tabular}
\end{table}

\subsubsection{Adhering to OMOP-CDM standards}
We prioritised data fidelity when aligning CEDRIC with OMOP-CDM, while conforming to standards wherever feasible. CEDRIC includes Australian-specific vocabularies, ACHI, ICD10-AM, and SNOMED CT-AU, which are not yet part of the standard OMOP-CDM concept set. National efforts are underway to integrate ACHI and ICD10-AM into OMOP vocabularies, after which full harmonisation of this dataset will be possible. In the interim, we incorporated these codes as custom concepts by loading CSV extracts into the \texttt{CONCEPT} table, assigning \texttt{concept\_id}s above two billion to avoid conflicts with standardised vocabularies \cite{carlson2024whowantstobea2billionaire}.

Differences in nullability rules also required design decisions. OMOP-CDM enforces many NOT NULL fields, whereas CEDRIC permits missing values broadly. We considered a YAML-based approach using a \texttt{non\_nullable\_cols} list to insert default values during loading, but ultimately modified the OMOP-CDM DDL to allow nulls in any non–primary key column. This ensured all source information could be retained—even where attributes were incomplete—reflecting our overall design goal of prioritising data completeness and source fidelity as part of an exploratory step towards a future anonymised and fully standardised dataset.

Table~\ref{tab:source_vocabs} summarises the major OMOP-CDM domains used and the corresponding CEDRIC vocabularies, ranging from nationally standardised terminologies to locally defined value sets.

\begin{table}[!t]
\centering
\caption{Source Vocabularies for OMOP-CDM Concept Domains}
\small
\setlength{\tabcolsep}{1mm}
\begin{tabular}{|
    m{36mm}|
    >{\raggedleft\arraybackslash}m{27mm}|
}
\hline
\textbf{OMOP Domain} & \textbf{Vocabulary} \\
\hline
Visit           & Local concept sets  \\
Condition       & ICD10AM  \\
Procedure       & ACHI  \\
Observation     & Local concept sets  \\
Measurement     & Local concept sets  \\
Note            & Local concept sets  \\
\hline
\end{tabular}
\label{tab:source_vocabs}
\end{table}

\section{Results}
\label{sec:results}
\subsection{Data Quality Assessment Using Data Quality Dashboard}
To verify data integrity and examine OMOP-CDM conformance, we ran the OHDSI Data Quality Dashboard (DQD) on the transformed dataset. Table~\ref{tab:dqd_results} shows an overall 97\% pass rate. Remaining failures reflect known characteristics and design choices:

\begin{itemize}
    \item \textbf{Plausibility}: Historical records predating 1950 trigger expected failures.
    \item \textbf{Conformance}: Limited relaxations of \texttt{NOT NULL} constraints were applied to preserve source nullability, supporting our goal of prioritising data completeness.
    \item \textbf{Completeness}: Some source values could not yet be mapped to standard OMOP concepts, consistent with an exploratory step toward a future anonymised and fully standardised dataset.
\end{itemize}

\begin{table}[!t]
\caption{Data Quality Dashboard Results Overview}
\small
\setlength{\tabcolsep}{2pt} 
\renewcommand{\arraystretch}{1.2} 
\begin{tabular}{|l|rrr|rrr|rrr|}
\hline
 & 
\multicolumn{3}{c|}{\textbf{Verification}} &
\multicolumn{3}{c|}{\textbf{Validation}} &
\multicolumn{3}{c|}{\textbf{Total}} \\
\cline{2-10}
 & \textbf{Pass} & \textbf{Fail} & \textbf{Pass} 
 & \textbf{Pass} & \textbf{Fail} & \textbf{Pass} 
 & \textbf{Pass} & \textbf{Fail} & \textbf{Pass} \\
\hline
Plausibility & 475 & 47 & 91\% & 291 & 0 & 100\% & 766 & 47 & 94\% \\
Conformance  & 898 & 1 & 100\% & 128 & 13 & 91\% & 1026 & 14 & 99\% \\
Completeness & 434 & 18 & 96\% & 16 & 1 & 94\% & 450 & 19 & 96\% \\
\hline
\textbf{Total} & 1807 & 66 & 96\% & 435 & 14 & 97\% & 2242 & 80 & \textbf{97\%} \\
\hline
\end{tabular}
\label{tab:dqd_results}
\end{table}

\subsection{Volume and Performance Metrics}
The ETL process successfully loaded a substantial portion of CEDRIC into OMOP-CDM. To evaluate pipeline efficiency, we collected metrics including:

\begin{itemize}
    \item \textbf{Rows inserted per table}: Core tables (e.g., \texttt{person}, \texttt{observation}, \texttt{condition\_occurrence}) were populated according to the original CEDRIC distributions.
    \item \textbf{Mapping table construction time}: Includes building internal lookup tables and applying YAML-defined concept mappings.
    \item \textbf{Data insertion time}: Time to transform and load source records into OMOP-CDM tables.
    \item \textbf{Total processing time per table}: Aggregated per-table ETL duration to inform potential optimisations.
\end{itemize}

Table~\ref{tab:etl_performance} summarises row counts and timing metrics across primary OMOP-CDM tables. Results indicate that the ETL framework efficiently handles datasets of varying size and complexity without modification to the core engine.

\begin{table}[!t]
\caption{ETL Performance Metrics for OMOP-CDM Tables}
\small
\setlength{\tabcolsep}{1mm}
\begin{tabular}{|
    m{21mm}|
    >{\raggedleft\arraybackslash}m{16mm}|
    >{\raggedleft\arraybackslash}m{12mm}|
    >{\raggedleft\arraybackslash}m{14mm}|
    >{\raggedleft\arraybackslash}m{14mm}|
}
\hline
\textbf{OMOP-CDM Table} & 
\textbf{Num Rows} & 
\textbf{Num Update Cols} &
\textbf{Mapping Insert Time (ms)} & 
\textbf{Total Time (ms)} \\
\hline
CARE\_\allowbreak SITE & 8,030 & 2 & 32 & 254 \\
& & & & \\
PROVIDER & 288 & 1 & 14 & 126 \\
& & & & \\
LOCATION & 2,759,586 & 8 & 35,875 & 274,299 (4.6 min) \\
& & & & \\
PERSON & 2,782,378 & 11 & 15,304 & 564,560 (9.4 min)\\
& & & & \\
VISIT\_\allowbreak OCCURRENCE & 27,250,260 & 12 & 79,286 (1.3 min) & 4,682,856 (1.3 hrs)\\
& & & & \\
VISIT\_\allowbreak DETAIL & 65,550,939 & 11 (22)\textsuperscript{*} & 732,529 (12.2 min) & 18,743,517 (5.2 hrs)\\
& & & & \\
CONDITION\_\allowbreak OCCURRENCE & 20,819,579 & 11 & 63,056 (1.1 min) & 2,523,176 (42 min)\\
& & & & \\
PROCEDURE\_\allowbreak OCCURRENCE & 6,821,962 & 9 & 15,400 & 659,267 (11 min)\\
& & & & \\
NOTE & 96,796,791 & 11 & 2,078,671 (34.6 min) & 41,215,966 (11.4 hrs)\\
& & & & \\
OBSERVATION & 8,110,299 & 8 & 303,528 (5.1 min) & 1,415,341 (23.6 min)\\
& & & & \\
MEASUREM-\allowbreak ENT & 54,508,593 & 8 & & $\mathtt{\sim}$3×10\textsuperscript{8} (3.5 days) \\ 

\hline
\multicolumn{5}{p{88mm}}{*The \texttt{VISIT\_DETAIL} ETL consisted of loads from two separate tables, both applying updates to the same 11 columns}
\end{tabular}
\label{tab:etl_performance}
\end{table}

\section{Findings}
\label{sec:findings}
We developed an end-to-end ETL pipeline to transform heterogeneous health data into OMOP-CDM. Starting with the \texttt{PERSON} table, nine destination columns were mapped from a CEDRIC patient view, and the ETL was validated iteratively on subsets from 100 to 25 million rows to assess scalability. Duplicate patients were resolved via a de-duplicated \texttt{SELECT DISTINCT} view. The mapping table captured CEDRIC-to-OMOP primary key relationships and was reused to maintain consistent foreign keys in \texttt{VISIT\_OCCURRENCE}, enforcing table execution order.

For the \texttt{MEASUREMENT} table, MongoDB sources were flattened into a staging table using YAML-defined rules and MongoDB aggregation pipelines. Batched processing leveraged the mapping table’s boolean \texttt{processed} flag, avoiding memory-intensive transformations in Python. Subsequent ETL steps proceeded on the staged data without modifying the YAML or ETL code.

Other tables—including \texttt{VISIT\_OCCURRENCE}, \texttt{PROCEDURE\_OCCURRENCE}, \texttt{CONDITION\_OCCURRENCE}, \texttt{MEASUREMENT}, \texttt{LOCATION}, \texttt{PROVIDER}, \texttt{CARE\_SITE}, and \texttt{NOTE}—were processed sequentially. \texttt{PROVIDER} records were simplified to role/specialty-based entries. After populating \texttt{CARE\_SITE} and \texttt{PROVIDER}, prior tables were updated with foreign keys (\texttt{care\_site\_id}, \texttt{provider\_id}) by re-running the ETL with \texttt{processed = 0}. \texttt{NOTE} entries store HTTPS links to the SWSLHD document system rather than text, offloading storage and access control to existing infrastructure.

\section{Discussion and Limitations}
\label{sec:discussion}
\subsection{Discussion}
This study evaluated the feasibility and effectiveness of a configurable ETL framework for transforming healthcare datasets into OMOP-CDM, addressing three key research questions on feasibility, automation and scalability, and development constraints without direct data access. Our results show that these challenges can be met using a flexible, schema-driven ETL process.

\subsubsection{Feasibility of a configurable ETL pipeline (RQ1)}
We confirmed feasibility by successfully transforming the CEDRIC dataset using a Python-based, configurable ETL framework. All source-specific logic is externalized in human-readable YAML files, which define schema mappings, value transformations, and OMOP-target logic. This approach allows the same ETL engine to be applied to diverse healthcare datasets with minimal changes, demonstrating portability, reuse, and the ability to abstract transformations away from hardcoded scripts.

\subsubsection{End-to-end automation, incremental updates, and scalability (RQ2)}
Our ETL framework provides a semi-automated pipeline to transform source T-SQL and MongoDB data into a compliant OMOP-CDM instance. While we modified the OMOP-CDM DDL for specific source requirements, the framework also supports standard OMOP-CDM setups. It allows re-execution without resetting the target database, enabling incremental updates and continuous integration, which is crucial for evolving healthcare datasets. YAML-based configuration ensures fine-grained control over mappings and transformations, balancing source data fidelity with OMOP-CDM standardisation.

\subsubsection{Semi-structured data integration from MongoDB (RQ3)}
We addressed the challenge of converting hierarchical, document-based data into the relational OMOP-CDM by extending the ETL to flatten nested MongoDB documents. YAML-defined rules guided the unwinding of nested structures into SQL staging tables, temporarily storing rows while preserving keys linking sub-documents to parent records. This approach maintained hierarchical traceability within the relational schema.

\subsection{Limitations}
While the ETL approach met its primary goals, several limitations suggest opportunities for improvement.

First, requiring source and destination databases on the same server simplifies orchestration with SQLAlchemy ORM but limits portability. Future work could explore cross-instance deployments.

Second, mapping local codes to standard vocabularies remains manual, especially for pathology and document types with limited coverage. Semi-automated mapping, community-shared vocabularies, and expanded standards could reduce reliance on non-standard concepts and enhance semantic interoperability.

Third, large MongoDB collections created throughput bottlenecks. Batching and staging mitigated this, but further gains could come from query optimization, indexing, selective extraction, and incremental loading.

Finally, YAML-based configuration supports transparency and reuse, but complex transformations sometimes required embedded SQL via aliased subqueries, highlighting a trade-off between declarative simplicity and SQL expressiveness.

These constraints outline a roadmap for improvement: enhanced connectivity, vocabulary harmonization, and optimized document-store processing can further improve scalability, interoperability, and maintainability.

\section{Conclusion}
\label{sec:conclusion}
We present a validated, configurable, and scalable ETL framework for transforming heterogeneous health data into OMOP-CDM. By extending YAML-based schema mapping to support MongoDB document stores, the pipeline balances source data fidelity with OMOP-CDM conformance.

The framework offers three main contributions: (1) a schema-agnostic design separating mapping logic from code to enhance reusability; (2) an end-to-end semi-automated pipeline supporting incremental updates in dynamic clinical environments; and (3) a strategy for flattening semi-structured data into relational tables while preserving hierarchical information. Together, these address key barriers to scaling OMOP-CDM adoption.

This work provides a blueprint for large-scale OMOP-CDM implementation, with future efforts focused on automating concept alignment and optimizing performance for massive document collections in distributed infrastructures.

\appendices 

\onecolumn
\section{YAML Object Definitions}
\label{appendix:yaml}
\begin{longtable}{
    >{\raggedright\arraybackslash}p{30mm}
    >{\raggedright\arraybackslash}p{42mm}
    >{\raggedright\arraybackslash}p{20mm}
    >{\raggedright\arraybackslash}p{78mm}
    }
    \caption{The \texttt{mongodb} YAML object} \\
    \hline
    \textbf{Type} &
    \textbf{Description} &
    \textbf{MQL Equivalent} &
    \textbf{Example Definition} \\
    \hline

    \addlinespace
    \multicolumn{4}{c}{\textbf{Fields comprising the \texttt{mongodb} YAML object}} \\
    \addlinespace
    \hline

    \texttt{source} & 
    The MongoDB collection. &
    N/A &
    \begin{minipage}[t]{\linewidth}
    \vspace{-0.5\baselineskip}
    \begin{lstlisting}[basicstyle=\ttfamily\footnotesize,aboveskip=0pt,belowskip=0pt]
source:
database: cedric
collection: |
  observations
    \end{lstlisting}
    \end{minipage} \\
    \addlinespace

    \texttt{staging\_table} & 
    The fully qualified name of the staging table. &
    N/A &
    \begin{minipage}[t]{\linewidth}
    \vspace{-0.5\baselineskip}
    \begin{lstlisting}[basicstyle=\ttfamily\footnotesize,aboveskip=0pt,belowskip=0pt]
staging_table: |
  OMOP.dbo.OBSERVATION_DOCUMENT
    \end{lstlisting}
    \end{minipage} \\
    \addlinespace

    \texttt{document\_match} & 
    An MQL predicate object, structured as a YAML object to be applied as a filter before flattening. &
    \texttt{\$match} &
    \begin{minipage}[t]{\linewidth}
    \vspace{-0.5\baselineskip}
    \begin{lstlisting}[basicstyle=\ttfamily\footnotesize,aboveskip=0pt,belowskip=0pt]
document_match:
  - document_status.form_date_time:
      $gte: 2024-01-01T00:00:00Z
  - form_name:
      $in:
        - "Vital Signs & Observations"
        - "Full Blood Count"
    \end{lstlisting}
    \end{minipage} \\
    \addlinespace

    \texttt{row\_match} & 
    An MQL predicate object, structured as a YAML object to be applied as a filter after flattening. &
    \texttt{\$match} &
    \begin{minipage}[t]{\linewidth}
    \vspace{-0.5\baselineskip}
    \begin{lstlisting}[basicstyle=\ttfamily\footnotesize,aboveskip=0pt,belowskip=0pt]
row_match:
  event_name:
    $in:
      - "Pulse Rate"
      - "Oxygen Delivery"
      - "Systolic BP"
    \end{lstlisting}
    \end{minipage} \\
    \addlinespace

    \texttt{fields} &
    A list of document fields to be included as columns in the staging table. The contents can be specified with arbitrarily nested self-similar fields objects, to be unwound into the flattened rows. &
    \texttt{\$unwind}, \texttt{\$unionWith}, \texttt{\$project} &
    \begin{minipage}[t]{\linewidth}
    \vspace{-0.5\baselineskip}
    \begin{lstlisting}[basicstyle=\ttfamily\footnotesize,aboveskip=0pt,belowskip=0pt]
fields:
  - doc_key_path: encounter_id
    staging_column: encounter_id
    datatype: INTEGER

  - doc_key_path: obs
    fields:
      - doc_key_path: event_name
        staging_column: event_name
        datatype: VARCHAR
    \end{lstlisting}
    \end{minipage} \\
    
    \hline

    \addlinespace
    \multicolumn{4}{c}{\textbf{Fields comprising the fields object within the \texttt{mongodb} YAML object}} \\
    \addlinespace
    \hline

\texttt{doc\_key\_path} & 
    The key within the MongoDB document, using a combination of dot notation for fields and index notation for array elements. &
    A field key, \texttt{\$getField}, \texttt{\$arrayElemAt} &
    \begin{minipage}[t]{\linewidth}
    \vspace{-0.5\baselineskip}
    \begin{lstlisting}[basicstyle=\ttfamily\footnotesize,aboveskip=0pt,belowskip=0pt]
doc_key_path: |
  update_history[0].activity_date_time
    \end{lstlisting}
    \end{minipage} \\
    \addlinespace

    \texttt{staging\_column} & 
    The name of the corresponding column in the staging table. &
    N/A &
    \begin{minipage}[t]{\linewidth}
    \vspace{-0.5\baselineskip}
    \begin{lstlisting}[basicstyle=\ttfamily\footnotesize,aboveskip=0pt,belowskip=0pt]
staging_column: activity_date_time
    \end{lstlisting}
    \end{minipage} \\
    \addlinespace

    \texttt{datatype} & 
    The SQL datatype of the corresponding staging column. &
    N/A &
    \begin{minipage}[t]{\linewidth}
    \vspace{-0.5\baselineskip}
    \begin{lstlisting}[basicstyle=\ttfamily\footnotesize,aboveskip=0pt,belowskip=0pt]
datatype: INTEGER
    \end{lstlisting}
    \end{minipage} \\
    \addlinespace

    \texttt{fields} & 
    A self-similar fields object capable of arbitrary nesting. &
    N/A &
    \begin{minipage}[t]{\linewidth}
    \vspace{-0.5\baselineskip}
    \begin{lstlisting}[basicstyle=\ttfamily\footnotesize,aboveskip=0pt,belowskip=0pt]
fields:
  - doc_key_path: clinical_event_id
    staging_column: event_id
    datatype: INTEGER

  - doc_key_path: measurements
    fields:
      - doc_key_path: clinical_event_id
        staging_column: event_id
        datatype: INTEGER
    \end{lstlisting}
    \end{minipage} \\
    \addlinespace
    
    \hline

\label{tab:mongo_yaml}
\end{longtable}    
\twocolumn

\onecolumn
\section{MongoDB Aggregation Pipeline}
\begin{figure}[H]
\begin{lstlisting}[frame=single,basicstyle=\ttfamily\small,breaklines=true]
[
    {$unwind: {path: "$measures"}},
    {
        $project: {
            measure_name: {$ifNull: ["$measures.name", null]}
        }
    },
    {
        $unionWith: {
            coll: "measures",
            pipeline: [
                {$unwind: {path: "$panels"}},
                {$unwind: {path: "$panels.measures"}},
                {
                    $project: {
                        panel_measure_name: {
                            $ifNull: [
                                "$panels.measures.name",
                                null,
                            ]
                        },
                        panel_name: {
                            $ifNull: ["$panels.name", null]
                        },
                    }
                },
            ],
        }
    },
    {
        $lookup: {
            from: "measures",
            localField: "_id",
            foreignField: "_id",
            as: "root",
        }
    },
    {
        $project: {
            panel_name: {$ifNull: ["$panel_name", null]},
            panel_measure_name: {
                $ifNull: ["$panel_measure_name", null]
            },
            visit_id: {
                $ifNull: [
                    {
                        $getField: {
                            field: "visit_id",
                            input: {$arrayElemAt: ["$root", 0]},
                        }
                    },
                    null,
                ]
            },
            measure_name: {$ifNull: ["$measure_name", null]},
        }
    },
]
\end{lstlisting}
\caption{MQL Aggregation Pipeline generated from the example YAML in Fig.~\ref{fig:mongodb_example_yaml}}
\label{fig:mongodb_example_pipeline}
\end{figure}
\twocolumn

\section*{Acknowledgment}

The authors acknowledge that generative AI tools were used to assist in editing of the manuscript. All substantive technical decisions and writing were made by the authors.

\section*{References}

\bibliographystyle{IEEEtran}
\bibliography{references}

\end{document}